\documentclass[twocolumn,showpacs,superscriptaddress,amsmath,amssymb,prl]{revtex4}

\usepackage{graphicx}%
\usepackage{dcolumn}%

\begin{document}

\title{Self-consistent self-energy analysis of photoemission data}

\author{A. A. Kordyuk}
\affiliation{Institute for Solid State Research, IFW-Dresden, Helmholtzstr.~20, D-01069 Dresden, Germany}
\affiliation{Institute of Metal Physics of National Academy of Sciences of Ukraine, 03142 Kyiv, Ukraine}

\author{S. V. Borisenko}
\author{A. Koitzsch}
\author{J. Fink}
\author{M. Knupfer}
\affiliation{Institute for Solid State Research, IFW-Dresden, Helmholtzstr.~20, D-01069 Dresden, Germany}

\author{H. Berger}
\affiliation{Institute of Physics of Complex Matter, EPFL, CH-1015 Lausanne, Switzerland}

\date{August 2, 2004}%

\begin{abstract}
Here we present the details of a self-consistent procedure of the photoemission data analysis within the self-energy approach introduced in Ref.~\onlinecite{Bare} (cond-mat/0405696). We derive the relations of the quasiparticle self-energy with the parameters provided by photoemission spectra, demonstrate self-consistency of the Kramers-Kronig procedure and its robustness in determination of the bare dispersion, examine the possible influence of a particle-hole asymmetry, discuss the necessity of a clear definition of the "kink", and demonstrate the applicability of the developed approach for a couple of samples. 
\end{abstract}


\maketitle

\section{Introduction}

Considerably improved characteristics of the electron analysers used in modern angle-resolved photoemission spectroscopy (ARPES) together with the accumulated experience in photoemission on superconducting cuprates \cite{DamascelliRMP03} as well as recent success in clarification of the underlying electronic structure \cite{FengPRL01, ChuangPRL01, KordyukPRL2002, NBS} have led to understanding that a careful analysis of the lineshape of photoemission spectra \cite{EschrigPRB03, ChubukovKink} is needed to clarify the origin of the main interaction which is responsible for the pairing mechanism. The nodal direction in the Brillouin zone (BZ) of superconducting cuprates, being responsible for their conducting properties, is of a special importance. While the pronounced doping dependence of the quasiparticle spectral weight in the antinodal region of the BZ unambiguously points to a magnetic origin of the strong electron-boson coupling seen by ARPES \cite{BorisenkoPRL2003, KimPRL2003, GromkoPRB2003}, the nodal direction, where the famous "kink" on the renormalized dispersion \cite{VallaSci99, BogdanovPRL00, KaminskiPRL01} has been shown to persists in the whole doping range \cite{LanzaraNature01, ZhouNature03}, is a subject of unresolved controversy between phonon \cite{LanzaraNature01, ZhouNature03} and spin-fluctuation \cite{KordyukPRL2004} scenarios. 

Recently we have shown that the spectral weight of electronic excitations detected by ARPES along the nodal direction can be self-consistently described within the quasiparticle self-energy approach \cite{Bare}. With the suggested procedure the real and imaginary parts of the self-energy in a wide frequency range as well as the bare band dispersion can be uniquely determined. We have shown that the procedure is applicable to an underdoped Bi-2212 in the pseudo-gap state and determined a number of correlation parameters for that sample.

Here we give a detailed description of the procedure, discussing its limitations and physical consequences. We derive the relations of the quasiparticle self-energy with the parameters of photoemission spectra, demonstrate self-consistency of the procedure and its robustness in determination of the bare band dispersion, examine the possible influence of a particle-hole asymmetry, discuss a necessity of clear definition of the "kink", and demonstrate the applicability of the developed approach for other samples: an overdoped Bi(Pb)-2212 and optimally doped Bi(La)-2201.

\section{Self-energy approach}

\begin{figure*}[t]
\includegraphics[width=17cm]{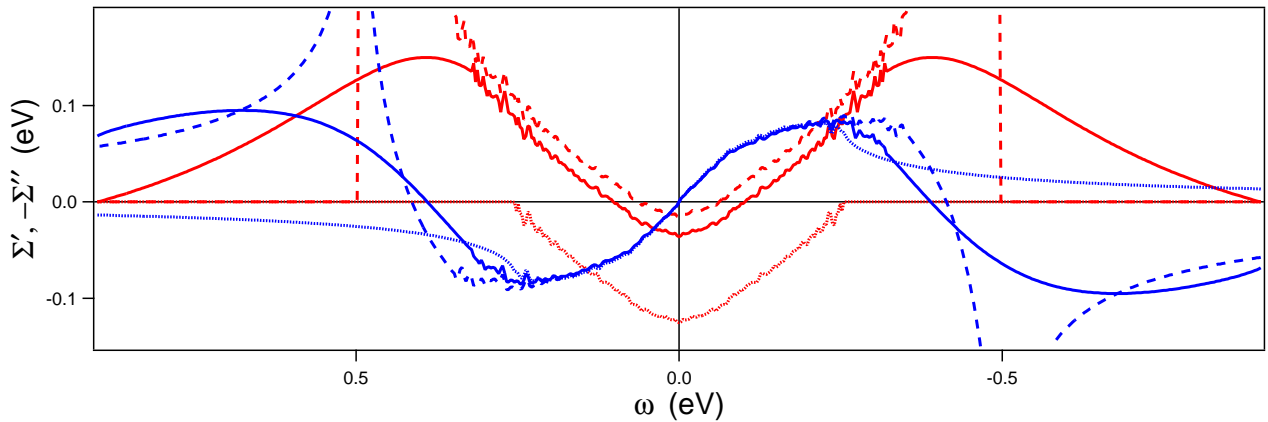}%
\caption{\label{Tails} Real (blue) and imaginary (red) parts of the self-energy related by Kramers-Kronig (KK) transform: $\Sigma' = \textbf{KK} \Sigma''$, for three models of $\Sigma''$ tails.}
\end{figure*}

We start with a definition of the spectral function within the self-energy approach:  
\begin{eqnarray}\label{A}
A(\omega, \mathbf{k}) = -\frac{1}{\pi}\frac{\Sigma''(\omega)}{(\omega - \varepsilon(\mathbf{k}) - \Sigma'(\omega))^2 + \Sigma''(\omega)^2}.
\end{eqnarray}
Here $\varepsilon(\mathbf{k})$ is the bare band dispersion and $\Sigma = \Sigma' + i\Sigma''$ is a quasiparticle self-energy, an analytical function the real and imaginary parts of which are related by the Kramers-Kronig (KK) transformation \cite{Landau}:
\begin{eqnarray}
\label{KK}\Sigma'(\omega) = \frac{1}{\pi}\,\,PV\int_{-\infty}^\infty{\frac{\Sigma''(x)}{x - \omega}\,dx},\\ 
\label{KK2}\Sigma''(\omega) = -\frac{1}{\pi}\,\,PV\int_{-\infty}^\infty{\frac{\Sigma'(x)}{x - \omega}\,dx}, 
\end{eqnarray}
where $PV$ denotes the Cauchy principal value. Within such a definition, $\Sigma''(\omega) < 0$, and $\Sigma'(\omega) > 0$ for $\omega < 0$.  

It is instructive to express some interaction parameters via both self-energy functions. For example, the coupling strength $\lambda$, which can be defined as
\begin{eqnarray}\label{lambda}
\lambda = - \left({\frac{d\Sigma'}{d\omega}}\right)_{\omega=0}
\end{eqnarray}
can be expressed in terms of $\Sigma''$ differentiating the KK relation (\ref{KK}):
\begin{eqnarray}\label{dKK}
\frac{d\Sigma'(\omega)}{d\omega} = \frac{1}{\pi}\,\,PV\int_{-\infty}^\infty{\frac{\Sigma''(x)-\Sigma''(\omega)}{(x - \omega)^2}\,dx}.
\end{eqnarray}
Here we used the fact that adding some constant to $\Sigma''(x)$ in (\ref{KK}) does not change the result. Then, for an even $\Sigma''(\omega)$,
\begin{eqnarray}\label{lambda}
\lambda = \frac{-2}{\pi}\int_0^\infty{\frac{\Sigma''(\omega) - \Sigma''(0)}{\omega^2}\,d\omega}.
\end{eqnarray}

Eq.~(\ref{lambda}) gives a certain feeling how $\Sigma''(\omega)$ function influences $\lambda$. For example, for a simple case
\begin{eqnarray}\label{S1}
\Sigma''(\omega) = -
	\begin{cases}
	\alpha \omega^2 + C & \text{for $|\omega| < \omega_c$}, \\
	0 & \text{for $|\omega| > \omega_c$},
	\end{cases}
\end{eqnarray}
where $\omega_c > 0$ is an energy cut-off and $C \equiv -\Sigma''(0) > 0$ is an offset, Eq.~(\ref{lambda}) gives 
\begin{eqnarray}\label{lambda_sim}
\lambda = \frac{2}{\pi} \left(\alpha \, \omega_c - \frac{C}{\omega_c} \right) \approx \frac{2}{\pi} \, \alpha \, \omega_c
\end{eqnarray}
for $C \ll \omega_c$. As we show below, the different but reasonable tails of~$\Sigma''(\omega)$ (at $|\omega| > \omega_c$) almost do not effect the experimentally determined $\lambda/\alpha$ ratio, i.e.~influence of the high energy region on $\lambda$ can be described by only one parameter, $\omega_c$.

\section{Nodal spectra analysis}

\begin{figure*}[]
\includegraphics[width=5cm]{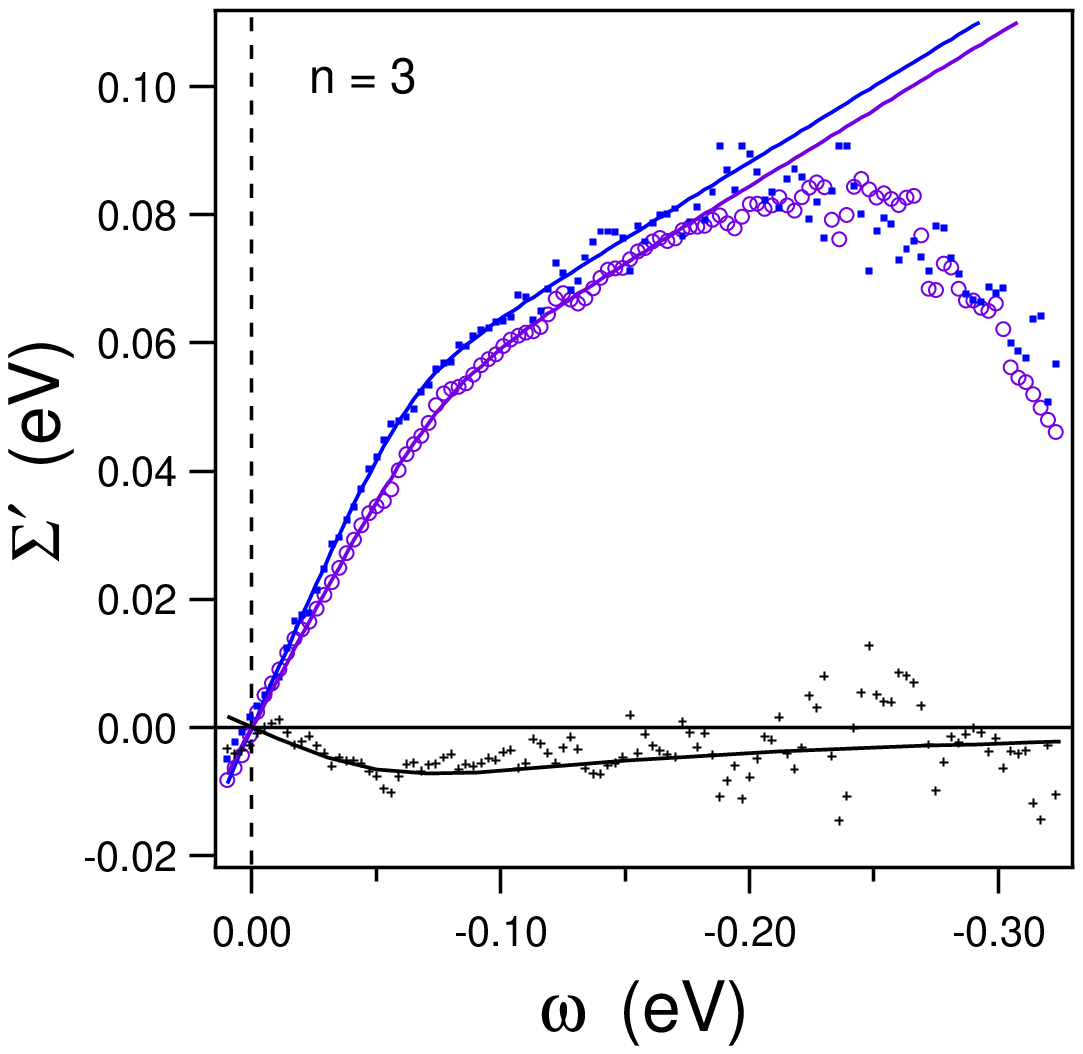}%
\includegraphics[width=5cm]{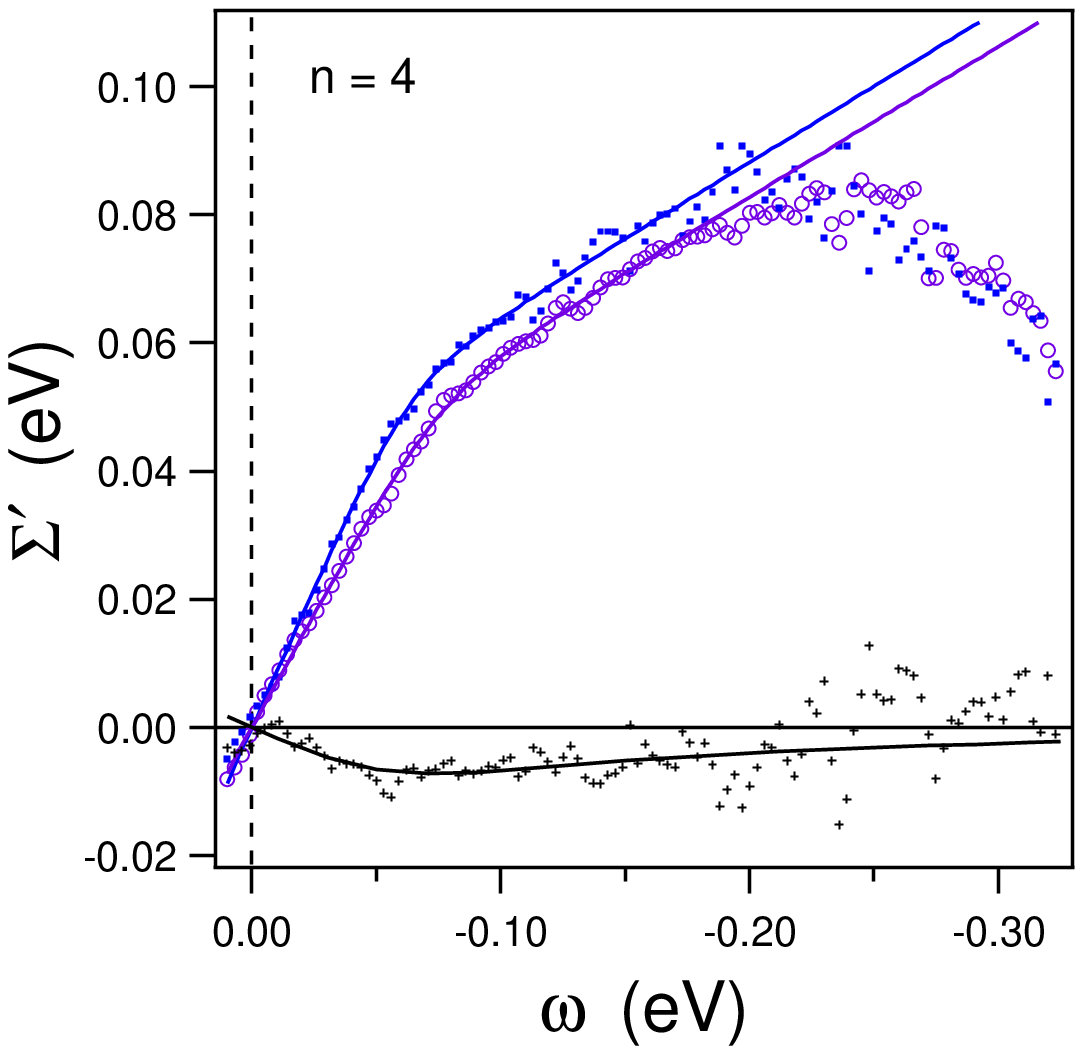}%
\includegraphics[width=5cm]{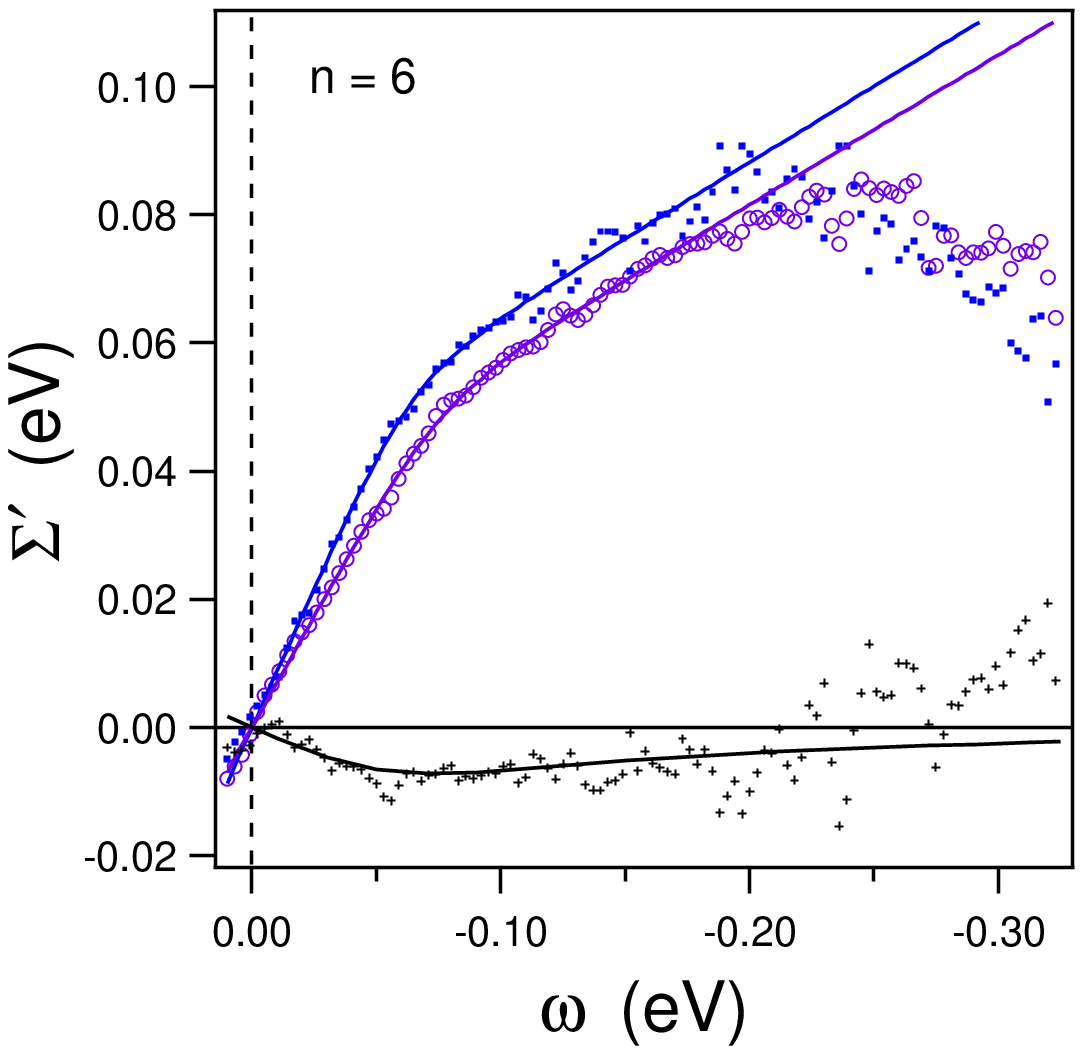}\\%
\includegraphics[width=5cm]{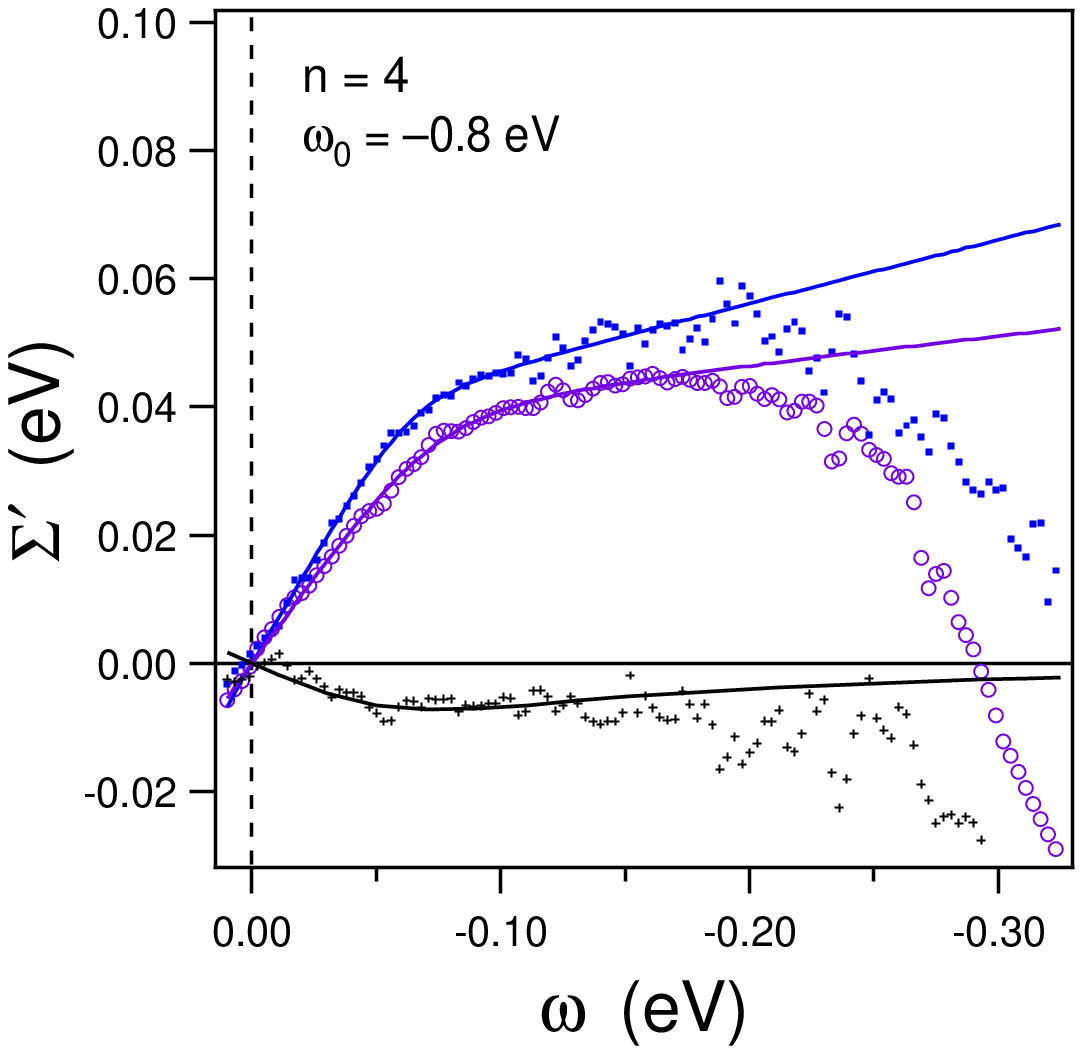}%
\includegraphics[width=5cm]{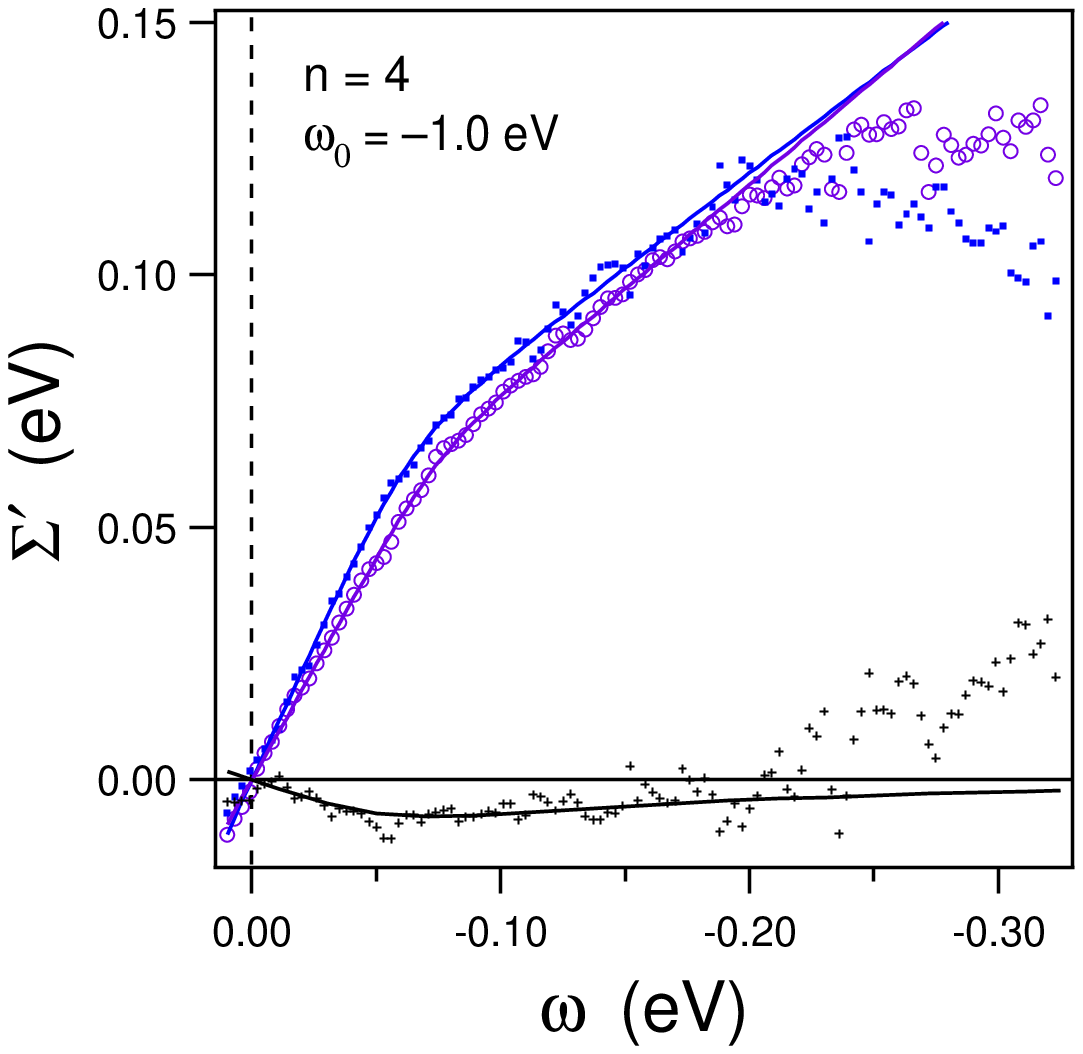}%
\caption{\label{ReSs} Illustration of the fitting procedure: real parts of the self-energy $\Sigma'_{disp}(\omega)$ (filled blue squares) obtained by (\ref{ReS1}) and $\Sigma'_{KK}(\omega)$ (open violet circles) by (\ref{ImS1}) and (\ref{KK}); the difference $\Delta\Sigma'(\omega) = \Sigma'_{KK}(\omega) - \Sigma'_{disp}(\omega)$ (small crosses) is fitted to $R'(\omega)$ (corresponding solid line), the contribution of overall resolution determined by (\ref{ImR})-(\ref{ReR}). In first three panels $\omega_0 = -0.9$ but different $n =$ 3, 4 and 6 in (\ref{S3}) are compensated by different $\omega_c = 0.34$, $0.45$ and $0.52$ eV respectively. Last two panels, the "best" fitting results for slightly different $\omega_0$'s.}
\end{figure*}

The self-energy approach can be applied to the nodal spectra analysis (e.g., see Refs.~\onlinecite{VallaSci99, BogdanovPRL00, KaminskiPRL01}) as far as the seed dispersion in (\ref{A}) along the nodal direction is identical to be the bare electron dispersion, not affected by either superconducting gap or pseudo-gap opening. 

Modern ARPES measures the photocurrent intensity of outgoing electrons, $I$, which is the quasiparticle spectral weight (the quasiparticle spectral function $A$ multiplied by the Fermi-function $f$) slightly modified by matrix elements, $M(\mathbf{k},h\nu)$, energy and momentum resolutions, $R_{\omega}(\omega)$ and $R_{k}(\mathbf{k})$, and an extrinsic background $B(\omega)$:
\begin{eqnarray}\label{I}
I(\omega,\mathbf{k}) \propto M A(\omega,\mathbf{k}) f(\omega) \otimes R_{\omega} \otimes R_{k} + B(\omega).
\end{eqnarray}
It is peculiar for all cuprates, and especially for Bi-cuprates, that the matrix elements variation with $k$ along the nodal direction is negligible \cite{BorisenkoPRB2001} and that the extrinsic background is $k$-independent \cite{KaminskiPRB2004}, thus can be easily subtracted. The energy and momentum resolutions produce rather small effect which we discuss later. So, the photocurrent intensity with subtracted background is simply proportional to the quasiparticle spectral weight, and the question is whether it can be completely described by Eq.~(\ref{A}).

The self-energy in Eq.~(\ref{A}) is believed to be momentum independent along the nodal direction, or at least much less dependent than on frequency: $d\Sigma/dk \ll v_F d\Sigma/d\omega$, where $v_F = (d\varepsilon / dk)_{k=k_F}$ is the bare Fermi velocity. This assumption is supported by a perfect Lorentzian lineshape of the momentum distribution curves (MDC's, MDC($k$) = $A(k)_{\omega = const}$) \cite{KaminskiPRL01}, although it has been pointed out \cite{RanderiaPRB2004} that a linear $k$-dependence of $\Sigma'$ is a more general case which leads to the Lorentzian lineshape of MDC's. We will discuss a possibile $k$-dependence of $\Sigma'$ below. 

As long as MDC's are Lorentzians, two quantities at each frequency can be derived from the experimental data: the MDC width, $W(\omega)$ (defined as the half width at half maximum), and the MDC peak position, $k_m(\omega)$, which is thus an inverse function to the renormalized dispersion. In the vicinity of the Fermi level one can consider the bare dispersion as linear, and the relations between the self-energy parts and these experimentally determined functions are especially simple \cite{Bare}:
\begin{eqnarray}
\label{ReS0} \Sigma'(\omega) &=& \omega - v_F [k_m(\omega)-k_F],\\
\label{ImS0} \Sigma''(\omega)&=& -v_F W(\omega).
\end{eqnarray}

Since $v_F$ is a priori unknown, the system (\ref{ReS0})-(\ref{ImS0}) is incomplete. One can complete the system by Eq.~(\ref{KK}) but encountering a problem of "tails", the high energy behavior of $\Sigma''(\omega)$ that is also unknown. For example, using another ultimate model for $\Sigma''$-tails,
\begin{eqnarray}\label{S2}
\Sigma''(\omega) = -
	\begin{cases}
	\alpha \omega^2 + C & \text{for $|\omega| < \omega_c$}, \\
	\alpha \omega_c^2 + C & \text{for $|\omega| > \omega_c$},
	\end{cases}
\end{eqnarray}
which approximates the saturation of scattering rate at high frequencies, one obtains $\lambda = 4\, \alpha \, \omega_c /\pi$ instead of (\ref{lambda_sim}).

In the next section we show that this problem can be partially solved considering a finite frequency range available from ARPES experiments. For this range, however, the deviation of the bare dispersion from linear becomes essential, therefore, we approximate it by a parabola $\varepsilon(k) = \omega_0(1-k^2/k_F^2)$, which fits well to the band dispersion in given direction obtained within the tight-binding model \cite{KordyukPRB2003}. Here $\omega_0 = -k_F v_F/2$ is the bottom of the bare band. Within this approximation, the Eqs.~(\ref{ReS0}) and (\ref{ImS0}) can be rewritten as:
\begin{eqnarray}
\label{ReS1}\Sigma'(\omega) &=& \omega - \omega_0\left[1-\frac{k_m^2(\omega)}{k_F^2}\right], \\
\label{ImS1}\Sigma''(\omega) &=& \frac{2\,\omega_0}{k_F^2}\, W(\omega) \sqrt{k_m^2(\omega)-W^2(\omega)}.
\end{eqnarray}

\section{Self-consistent KK-transform}

The fitting machinery is based on Eqs.~(\ref{ReS1}), (\ref{ImS1}) and (\ref{KK}). One can define three steps here. In the two first, the real part of the self-energy, for given $\omega_0$, $\omega_c$ and $n$ (which characterizes the tails, see below), is calculated in two ways: (i) $\Sigma'_{disp}$ by Eq.~(\ref{ReS1}); (ii) $\Sigma'_{KK}$ by Eq.~(\ref{ImS1}) with subsequent KK transform (\ref{KK}). Then, in step (iii), the parameters $\omega_0$, $\omega_c$ and $n$ are varied until $\Sigma'_{disp}(\omega)$ and $\Sigma'_{KK}(\omega)$ coincide. In practice, we fit the difference $\Sigma'_{disp}-\Sigma'_{KK}$ to a small contribution of experimental resolution.

The calculation of $\Sigma'_{KK}$ deserves to be considered in details. In order to perform a KK transform, high-energy tails should be attached to $\Sigma''(\omega)$ derived from Eq.~(\ref{ImS1}). Eqs.~(\ref{S1}) and (\ref{S2}) represent two extremes which can be enclosed in a simple analytical expression:
\begin{eqnarray}\label{S3}
\Sigma''_{mod}(\omega) = -\frac{\alpha \, \omega^2 + C}{1+\left|\frac{\omega}{\omega_c}\right|^n},
\end{eqnarray}
as the ultimate cases with $n \to \infty$ and $n = 2$ respectively. For given $n$ and $\omega_c$, we construct $\Sigma''(\omega)$ function in a wide frequency range (up to $|\omega_0|$ or higher) assuming the particle-hole symmetry:
\begin{eqnarray}\label{BigS}
\Sigma''(\omega) = 
	\begin{cases}
	\Sigma''_{width}(|\omega|) & \text{for $|\omega| < \omega_m$}, \\
	\Sigma''_{mod}(\omega) & \text{for $|\omega| > \omega_m$},
	\end{cases}
\end{eqnarray}
where $\omega_m$ is a "confidence limit", a maximal experimental binding energy to which both the $W(\omega)$ and $k_m(\omega)$ functions can be confidently determined, $\Sigma''_{width}(\omega)$ is calculated from Eqs.~(\ref{ImS1}) for given $\omega_0$, $\Sigma''_{mod}(\omega)$ is fitted to $\Sigma''_{width}(\omega)$ in the confidence range in order to define $\alpha$ and $C$. Then, we obtain $\Sigma'_{KK}(\omega)$ by KK transform (\ref{KK}). 

Fig.~\ref{Tails} shows the pairs of $\Sigma''(\omega)$ and $\Sigma'(\omega)$ functions obtained in such a way for the same $\omega_0$ but for three different models: (\ref{S1}), (\ref{S2}), and (\ref{S3}) with $n = 4$ (dashed, dotted, and solid lines respectively). Since $\mathbf{KK} C = 0$, in order to simplify numerical calculation, the offset of $\Sigma''(\omega)$ curves is set to $\Sigma''(\omega_0) = 0$. The experimental data are taken for an underdoped Bi(Pb)-2212 (UD77, $T_c =$ 77 K) at 130 K and the fitting result for which $\omega_0 = -0.9$ eV is represented by solid-line $\Sigma'(\omega)$ \cite{Bare}.

In step (iii), as we mentioned above, the difference $\Delta\Sigma'(\omega) = \Sigma'_{KK}(\omega) - \Sigma'_{disp}(\omega)$ should be fitted not to zero but to some small but detectable contribution of the overall resolution $R'(\omega)$ \cite{Bare}. This difference can be easily understood by reasoning that finite energy and angular resolutions mainly effect the MDC's width rather than its peak position and that its contribution is frequency dependent. In order to illustrate this we can take into account the overall resolution, $R$, as $\Sigma''_{width}(\omega) = \sqrt{\Sigma''(\omega)^2 + R^2}$. Then one can consider its frequency dependent contribution to the imaginary part of $\Sigma(\omega)$ as the difference between $\Sigma''_{width}(\omega)$ and real $\Sigma''(\omega)$:
\begin{eqnarray}
\label{ImR} R''(\omega) &=& \sqrt{R^2 + \Sigma''(\omega)^2} - \Sigma''(\omega),
\end{eqnarray}
and, due to additivity of the KK-transform, $\Sigma'_{KK} = \mathbf{KK}\Sigma''_{W} = \mathbf{KK}\Sigma'' + \mathbf{KK}R'' = \Sigma'_{disp} + R'$, construct $\omega$-dependent contribution to $\Sigma'_{KK}$ as
\begin{eqnarray}
\label{ReR}	R'(\omega) &=& \mathbf{KK} R''(\omega).
\end{eqnarray}

Although, in principle, the resolution effect $R'(\omega)$ can be explicitely calculated from known energy and momentum resolutions, here we derive it empirically using $R$ as a parameter. It is seen from Fig.~\ref{Tails} that different tails do not affect the energy region $|\omega| <$ 0.25 eV, so, an irreducible difference in the slopes (see Fig.~\ref{ReSs}) $\Delta = d\Sigma'_{KK}(\omega)/d\omega - d\Sigma'_{disp}(\omega)/d\omega > 0$ in the low energy range $|\omega| <$ 0.07 eV (while $\Delta = 0$ at higher energies 0.1 eV $< |\omega| <$ 0.2 eV) is a measure of $R'(\omega)$. 

In Fig.~\ref{ReSs} we plot $R'(\omega)$ setting the offset of $\Sigma''(\omega)$ to zero that gives the value of $R$ = 0.015 eV. For $\Sigma''(0) < 0$ the procedure gives larger $R$ values to accommodate the difference in slopes but this does not affect the fact that the irreducible difference between $\Sigma'_{KK}(\omega)$ and $\Sigma'_{disp}(\omega)$ is caused by the experimental resolution, and depends on frequency like it is shown in Fig.~\ref{ReSs}: it vanishes at zero and high frequencies having a maximum around 0.1 eV.

Thus, we can visualize the fitting procedure as fitting the difference $\Delta\Sigma'(\omega)$ to $R'(\omega)$ function. The procedure has appeared to be robust with respect to the $\omega_0$ determination. Fig.~\ref{ReSs} illustrates this. First three panels show that for a correct value of $\omega_0 = -0.9$ eV there is space for other parameters to vary: different tails can be compensated by different $\omega_c$'s, e.g., for $n =$ 3, 4 and 6 in Eq.~\ref{S3}, $\omega_c = 0.34$, 0.45 and 0.52 eV respectively. On the other hand, at slightly different $\omega_0$'s (about 10\% lower and higher, see two right panels), $\Delta\Sigma'(\omega)$ cannot be fitted to $R'(\omega)$ in the whole frequency range.

\section{Other examples and physical consequences}

\begin{figure*}[t]
\includegraphics[width=5.5cm]{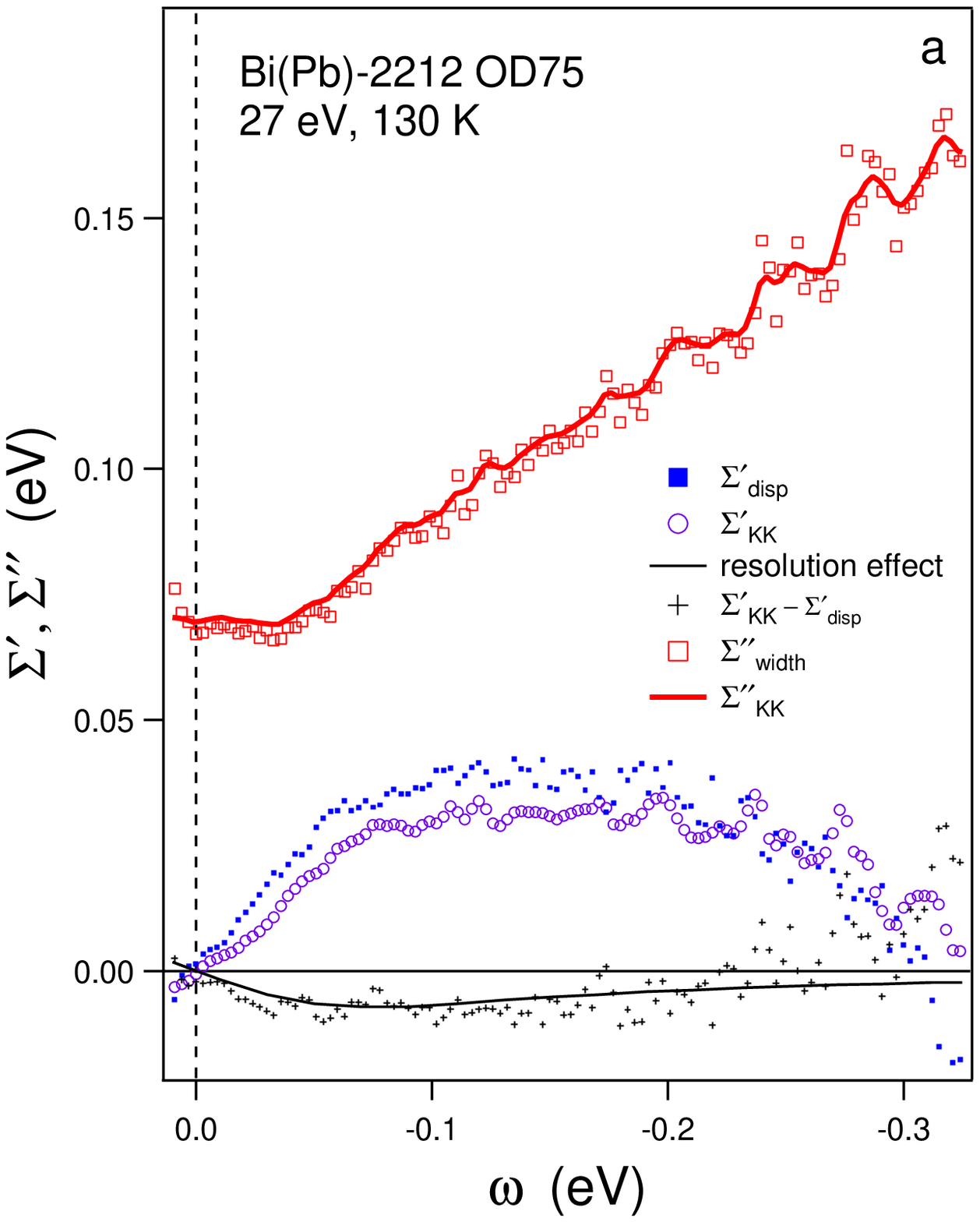}%
\includegraphics[width=5.5cm]{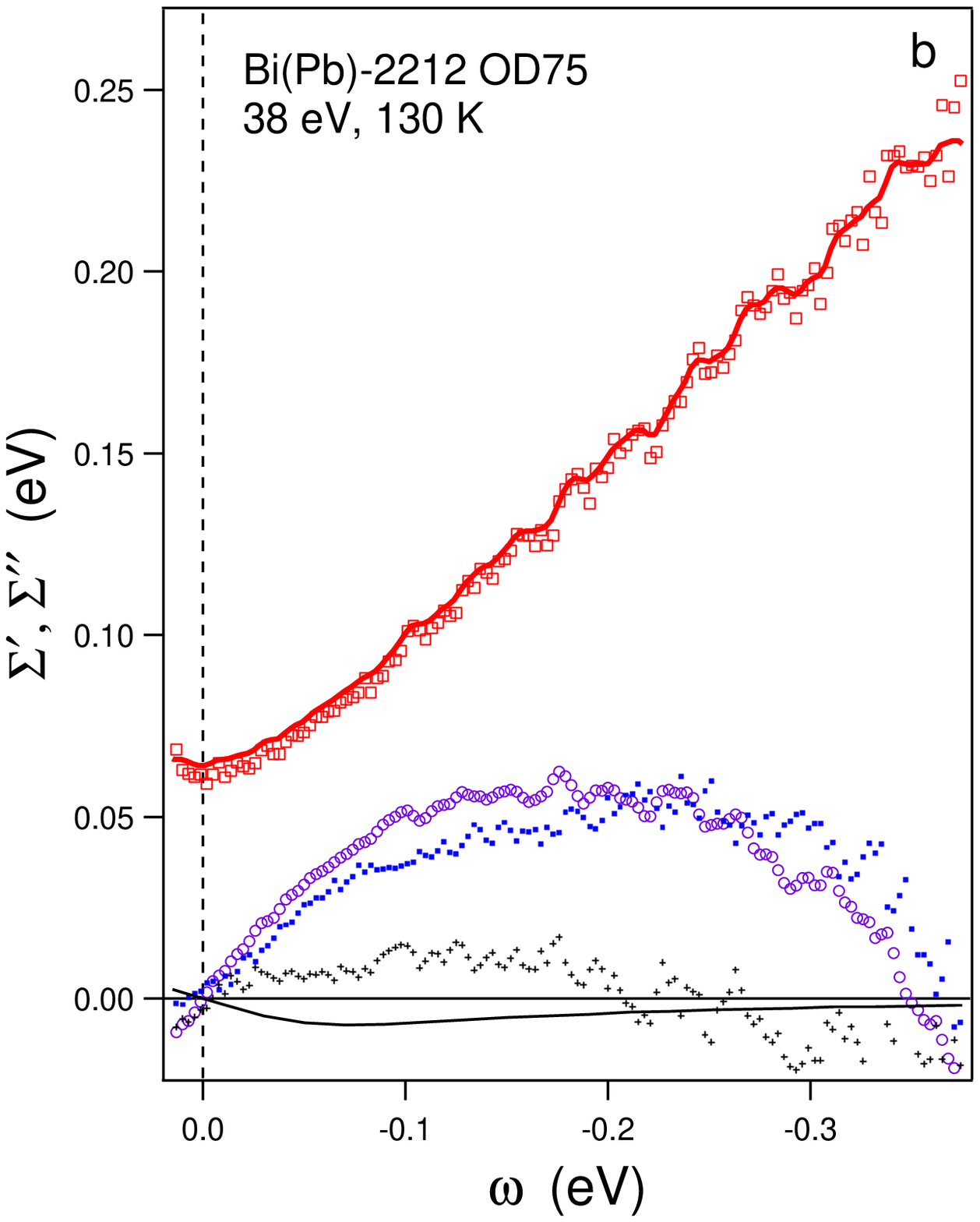}%
\includegraphics[width=5.5cm]{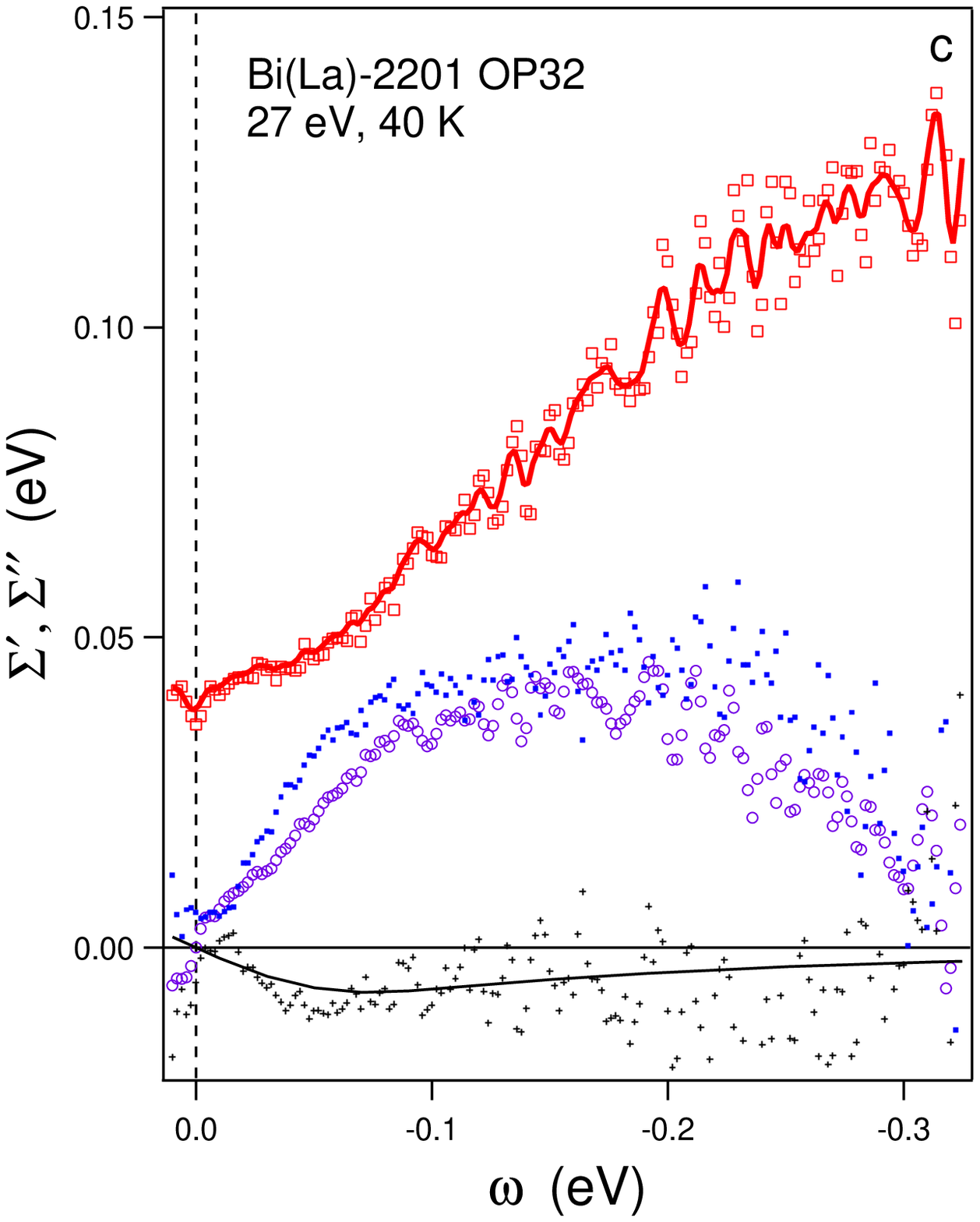}%
\caption{\label{Results} Real and imaginary parts of the self-energy extracted from the experiment with the described procedure.}
\end{figure*}

In Ref.~\onlinecite{Bare}, as mentioned above, we have applied the technique to underdoped Bi(Pb)-2212, UD77, in the pseudo-gap state (at 130 K). Here we present some other examples.

Fig.~\ref{Results}a shows the results for an overdoped Bi(Pb)-2212, OD75, at 130 K. Here, in order to check the correctness of the KK numerics, we also plot $\Sigma''_{KK}(\omega)$ function which is obtained by back KK-transform (\ref{KK2}) of $\Sigma'_{KK}(\omega)$. The parameters for this fit are: $\omega_0 = -0.86 \pm 0.03$ eV, $\omega_c = 0.40 \pm 0.05$ eV, $n = 4 \pm 0.5$. In case of the OD sample, the parameters $\omega_c$ and $n$ are better determined because of a higher confidence limit $\omega_m$ = 0.45 eV at which one can see that $\Sigma''(\omega)$ starts to saturate. We note that the lower value of $\omega_0$ compared to the underdoped sample is in agreement with the rigid band approximation.

The demonstrated self-consistency of the analyzed data shows, besides the applicability of the self-energy approach to superconducting cuprates, that the measured spectra belong to a single band and are free of influence of any additional features like other bands, superstructure or $k$-dependent background. It has been shown recently \cite{NBS} that although the electronic dispersion along the nodal direction in the bi-layer Bi-2212 is not degenerated, i.e.~has a finite splitting about 0.05 eV for the bare dispersion, the photoemission from the bonding band is highly suppressed at exactly 27 eV excitation energy. At other energies we do not expect that the described fitting procedure will work until the contributions of each band can be separated. Fig.~\ref{Results}b demonstrates this showing the "best" fitting result that can be achieved for $h\nu =$ 38 eV. The difference between $\Delta\Sigma'(\omega)$ and $R'(\omega)$ is apparent.

Fig.~\ref{Results}c shows another example of successful application of the described procedure to a single-layer Bi(La)-2201, OP34, at 40 K. In this case $\omega_0 = -0.79 \pm 0.05$ eV, $\omega_c = 0.37 \pm 0.07$ eV, $n = 3 \pm 0.6$.

\begin{figure}[b]
\includegraphics[width=7cm]{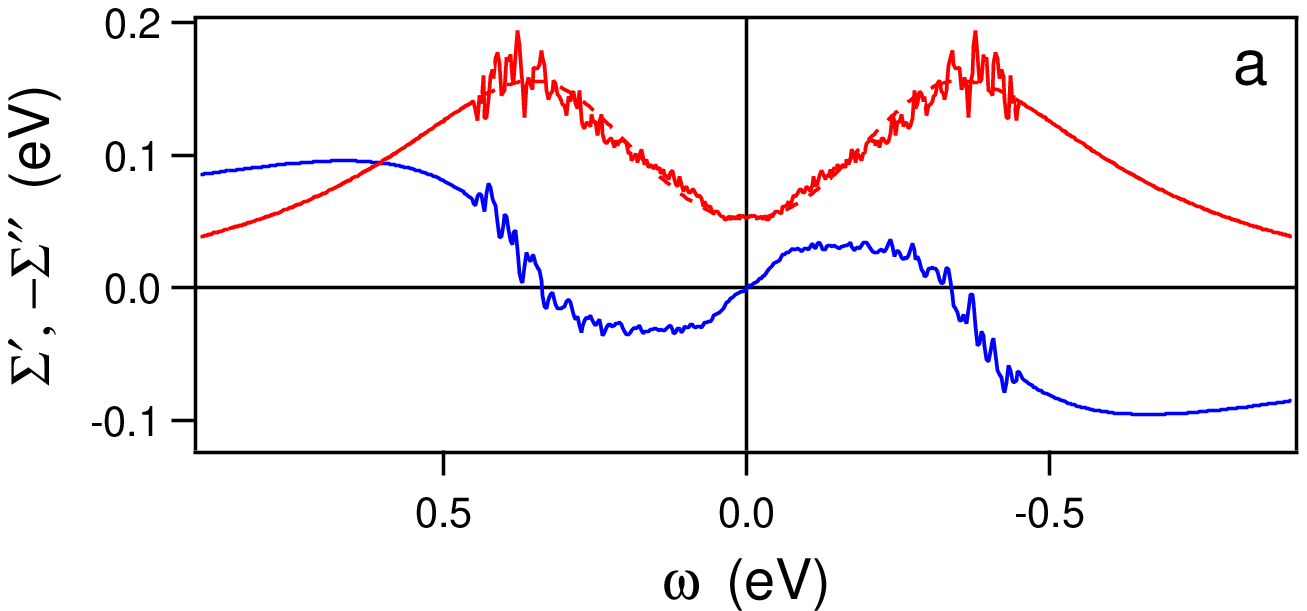}\\%
\includegraphics[width=7cm]{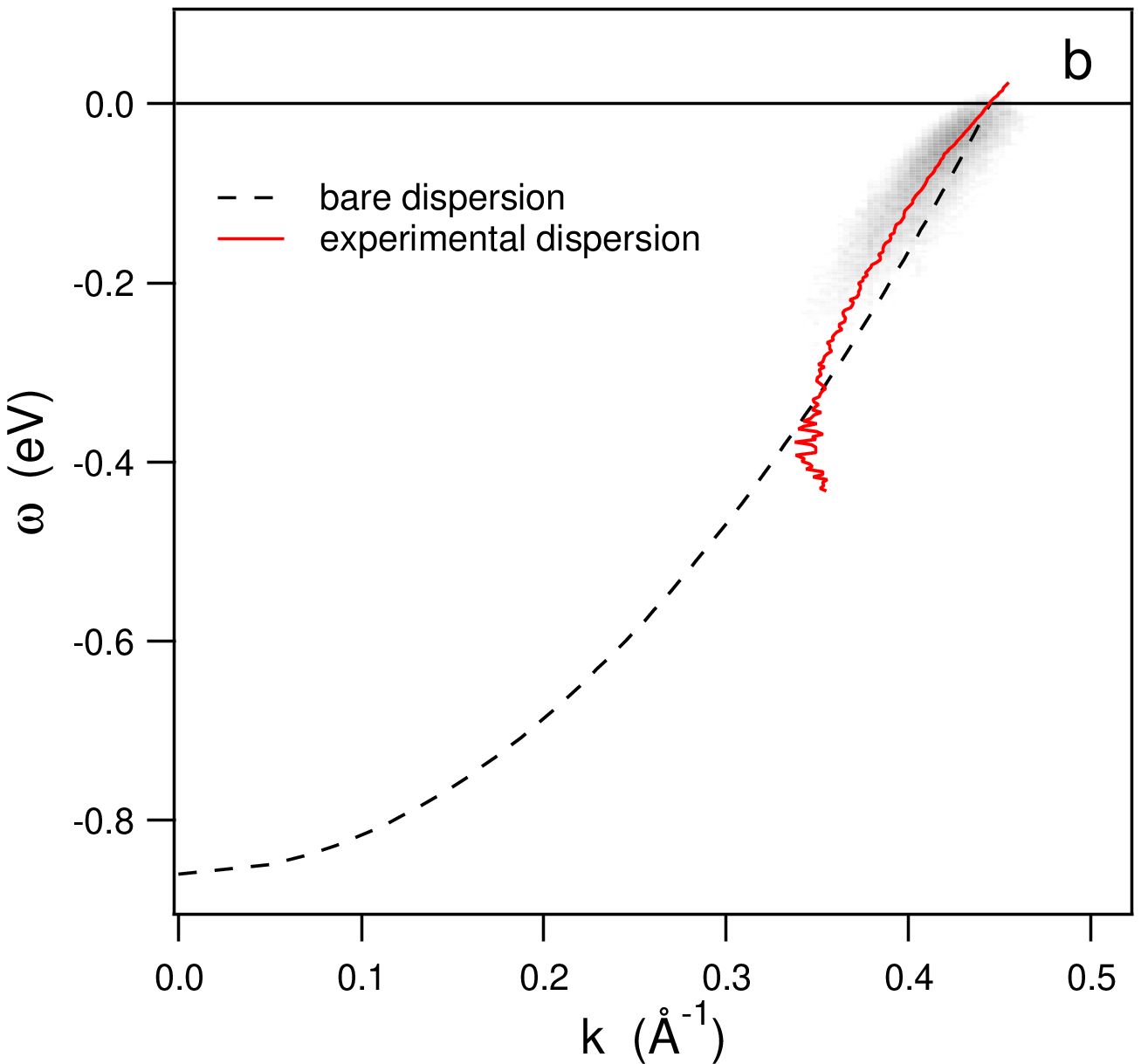}%
\caption{\label{OD75} The results of the fitting procedure for Bi(Pb)-2212 OD75: (a) real (blue line) and imaginary (red line) parts of the self-energy; (b) the experimental (solid red line) and bare (dashed line) dispersions on top of the experimentally measured quasiparticle spectral weight.}
\end{figure}

The presented examples purpose to illustrate the applicability of the self-energy approach to Bi-cuprates. We believe that the described procedure gives a powerful technique to purify the ARPES data from artificial features and to build a strong experimental basis for understanding of the nature of electronic interactions in cuprates, but still a big work on the data analysis should be performed. Nevertheless, some conclusions can be made even on this stage.  

It is interesting to note that even for UD77 sample, for which the saturation of $\Sigma''(\omega)$ has not been observed, it is not possible to reconcile the high-energy behavior of $\Sigma''(\omega)$ with the saturation extreme (\ref{S2}) [or (\ref{S3}) with $n = 2$]. This means that $|\Sigma''(\omega)|$ reaches the maximum and starts to decrease at about $\omega_c$, and, consequently, $\Sigma'(\omega)$ changes the sign at approximately the same frequency (see Fig.~\ref{Tails}). For OD and OP samples this conclusion is even more strict due to smaller bandwidth. Fig.~\ref{OD75} shows the results for Bi(Pb)-2212 OD75: (a) $\Sigma'(\omega)$ and $\Sigma''(\omega)$; (b) $k_m(\omega)$ and $\varepsilon(k)$ on top of the experimentally measured quasiparticle spectral weight.

The fact that $\omega_c$ is not equal but roughly two times less than $|\omega_0|$ is consistent with presence of an essential electron-electron scattering channel \cite{KordyukPRL2004} which mainly determines the lifetime of quasiparticles at high frequencies. Other consequences are discussed below.

\section{Model assumptions}

\begin{figure}[t]
\includegraphics[width=8cm]{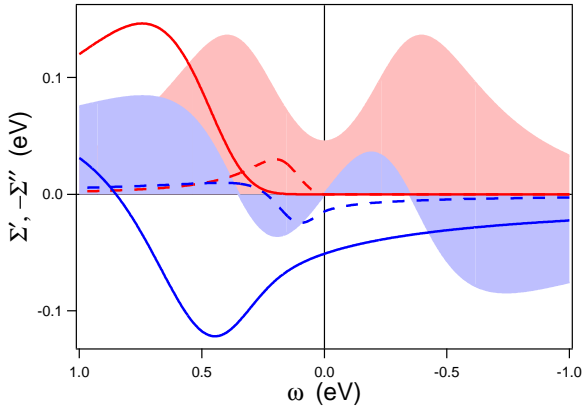}%
\caption{\label{asymmetry} Possible particle-hole asymmetry effect on $\Sigma''(\omega)$ (red) and $\Sigma'(\omega)$ (blue): low-energy (dashed lines, $\omega_{c}$ = 0.1 eV) and high-energy (solid lines, by Eq.~\ref{asymS}) contributions shown on the top of the symmetric self-energy (shaded areas).}
\end{figure}

First we focus on two assumptions which have been made about the model self-energy: $k$-independence and particle-hole symmetry.

It has been mentioned above that the symmetric Lorentzian lineshape of the MDC's taken along the nodal direction was considered as an experimental evidence that the quasi-particle self-energy hardly depends on momentum \cite{KaminskiPRL01}. Recently \cite{RanderiaPRB2004}, it has been noticed that the necessary condition for the Lorentzian lineshape is $\partial\Sigma''(k,\omega)/\partial k = 0$, but $\partial\Sigma'/\partial k$ can be an $\omega$-independent constant. This is especially interesting because the authors of Ref.~\onlinecite{RanderiaPRB2004} have shown that such a linear $k$-dependence of $\Sigma'$ can explain a non-trivial doping-dependent high-energy dispersion observed for a variety of cuprates \cite{ZhouNature03}.

As long as $\Sigma'(k,\omega) = \Sigma'_k(k) + \Sigma'_{\omega}(\omega)$ and $\partial\Sigma''/\partial k = 0$, $k$-dependence of $\Sigma'$ does not affect any result of the presented analysis except the bare dispersion. In this case, the real bare dispersion is just $\varepsilon^{real}(k) = \varepsilon(k) - \Sigma'_k(k)$, or $v_F^{real} = v_F - (\partial\Sigma'/\partial k)_{k = k_F}$. Although our preliminary results, being in agreement with band structure calculations \cite{NBS} and experimental plasmon dispersion \cite{NuckerPRB1991}, do not support strong $k$-dependence of $\Sigma'$, it will be interesting to examine its possible contribution in a wide doping range and for different compounds.

A possible particle-hole asymmetry is another complication which can effect the results of the presented analysis. In general, one can expect an asymmetry of the self-energy due to an asymmetric electron-boson interaction or as a simple consequence of asymmetric density of states. Without considering the origin of the asymmetry, we examine its possible influence based on the energy scale where it can appear. It is well known that because of the possibility to perform ARPES at finite temperature one can get the information about quasiparticle spectral weight not only below the chemical potential but also from some region above \cite{SatoPRB2001}. For $T$ = 300 K the MDC width can be measured up to 50 meV above $E_F$, and, within the experimental uncertainty, it has appeared to be completely symmetric (e.g., see \cite{KordyukPRL2004}). This means that if there is some asymmetry in the scattering rate at low energy scale ($\sim$ 0.1 eV, a characteristic scale which can originate from an electron-boson interaction or from the van Hove singularity in the occupied density of states of the hole-doped cuprates), its magnitude is too small to be seen in the $|\omega| <$ 50 meV energy range and, consequently, hardly effects the quasiparticle renormalization in the occupied region ($\omega < 0$). Fig.~\ref{asymmetry} illustrates this: the dashed curves, on top of the symmetric self-energy shown by shaded areas, represent a low-energy asymmetric contribution which is too big not to be noticed in $\Sigma''(\omega)$ (for $|\omega| <$ 50 meV) but too small to influence $\Sigma'(\omega < 0)$. The solid curves in Fig.~\ref{asymmetry} present the case of high-energy asymmetry that can steam from the asymmetry of the bare band \cite{KordyukPRB2003}. We simulate it by an asymmetry part in the scattering rate:
\begin{eqnarray}\label{asymS}
\Sigma_{a}''(\omega) = 
	\begin{cases}
	\Sigma_{mod}''(\omega,\omega_{c2})-\Sigma_{mod}''(\omega,\omega_{c}), & \omega > 0, \\
	0, & \omega < 0,
	\end{cases}
\end{eqnarray}
where $\Sigma_{mod}''$ is determined by (\ref{S3}) with $\omega_{c}$ = 0.45 eV, $\omega_{c2}$ = 0.66 eV, $n = 4$, $C = 0$. It is seen that although the influence of $\Sigma_{a}'(\omega)$ on renormalization at $-0.5$ eV $< \omega < 0$ eV is rather small (can be approximated at this stage by a linear contribution with a slope of about 20\% of $\lambda$) it can be, in principle, detected by more precise analysis, in which the influence of the energy and angular resolutions is taken into account explicitly.

\section{Phenomenology of the kink}

\begin{figure}[t]
\includegraphics[width=8cm]{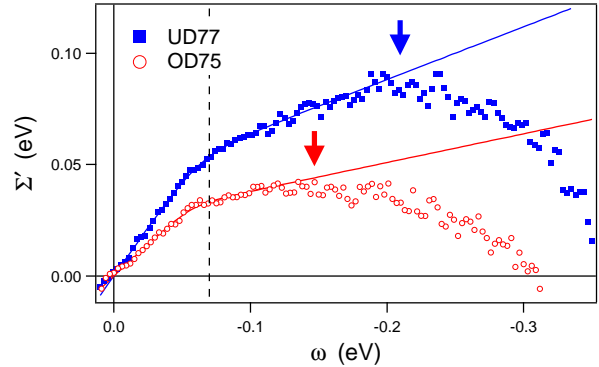}%
\caption{\label{kinks} The real parts of the self-energy for UD77 and OD75 samples at 130 K: solid lines show the result of fitting these real parts to (\ref{kink_function}) in a frequency range 0.17 eV $< \omega < 0$ for UD77 and 0.12 eV $< \omega < 0$ for OD75; looking down arrows mark $\Sigma'(\omega)$ maxima; the dashed line denotes the 70 meV "kink" energy.}
\end{figure}

In Fig.~\ref{kinks} we plot together the real parts of the self-energy for UD77 and OD75 samples at 130 K. Just from visual comparison of these data one can conclude that: (i) the renormalization for UD77 is considerably higher than for OD75; (ii) the energy of the maximum of $\Sigma'(\omega)$ for the overdoped sample is lower than for the underdoped sample, it is about two times closer to the 70 meV "kink" energy; (iii) the kink feature is well defined in the underdoped case and becomes weaker with overdoping. 

Following this tendency one can expect that with overdoping the 70 meV kink vanishes while the renormalization maximum moves to lower frequencies faking a persistence of the kink in the whole doping range. Therefore, it is clear that in order to clarify the origin of the kink feature a quantitative measure of it is required. 

Keeping the visual definition of the kink as a sharp bend of the renormalized dispersion, we formalized it in Ref.~\onlinecite{Bare} as a peak in the second derivative of $\Sigma'(\omega)$ and fitted it to a simple empirical function:
\begin{eqnarray}\label{kink_function}
\Sigma'_\textrm{low}(\omega) &=& -\lambda \omega - \frac{\Delta\lambda}{\pi} (\omega - \omega_k) \times \\ \nonumber
&&\times \left(\arctan \frac{\omega_k}{\delta} + \arctan \frac{\omega - \omega_k}{\delta} \right),
\end{eqnarray}
which gives a squared Lorentzian in a second derivative:
\begin{eqnarray}\label{ddkink_function}
K(\omega) = -\frac{d^2\Sigma'_\textrm{low}(\omega)}{d\omega^2} &=& \frac{2}{\pi}\frac{\delta^3 \Delta\lambda}{\left[\delta^2 + (\omega - \omega_k)^2\right]^2}.
\end{eqnarray}
Fitting $\Sigma'(\omega)$ of the underdoped sample in $|\omega| <$ 170 meV energy range to this formula we have obtained
an energy of the kink $\omega_k \approx -63$ meV, a kink width [half width at quarter maximum of $K(\omega)$] $\delta \approx$ 30 meV, and a strength of the kink $\int K d\omega = \Delta\lambda \approx 0.65$. For the overdoped sample $\omega_k \approx -56$ meV, $\Delta\lambda \approx 0.45$.  We believe that a systematic study of this or similar quantitaties as a function of doping and temperature will help to find the origin of the main electronic interaction in superconducting cuprates.

\section{Conclusions}
    
The self-energy approach is shown to be applicable to single band photoemission spectra for underdoped, overdoped and optimally doped Bi-based cuprates. The demonstrated self-consistency of the procedure opens a way to validate the photoemission spectra: the KK-sieve can be used to verify the spectra for the absence of the band splitting or artificial features. The preliminar analysis of the spectra certified in such a way shows that the overall renormalization as well as kink in the nodal direction of Bi-based cuprates is highly doping dependent, decreasing with overdoping. In the light of the present dilemma about the origin of the main scattering boson in the cuprates, a systematic quantitative analysis of the KK verified spectra measured at different temperature and doping level is indispensable.     

The project is part of the Forschergruppe FOR538 and is supported by the DFG under grants number KN393/4 and 436UKR17/10/04 and by the Swiss National Science Foundation and its NCCR Network "Materials with Novel Electronic Properties". We are grateful to S.~Ono and Yoichi Ando for supplying the Bi-2201 sample.

\end{document}